\documentclass[a4paper]{aa}
\usepackage{amsmath}
\usepackage{graphicx}
\usepackage{siunitx}
\usepackage{natbib}
\usepackage{multirow}
\usepackage{color}
\usepackage{array}
\usepackage{ifthen, hyperref}
\usepackage{fixltx2e}
\usepackage{lscape}
\usepackage{tabularx}
\usepackage{breakurl}
\bibpunct{(}{)}{;}{a}{}{,}

\begin{document}

\title{Constraining the role of novae as progenitors of Type Ia Supernovae}
\author{Monika~D.~Soraisam\thanks{\email{monikas@mpa-garching.mpg.de}} \inst{1}
    \and Marat~Gilfanov\thanks{\email{gilfanov@mpa-garching.mpg.de}} \inst{1,2,3}
    }

\institute{Max Planck Institute for Astrophysics, Karl-Schwarzschild-Str.~1, 85748 Garching, Germany\label{inst:MPA}
    \and Space Research Institute, Russian Academy of Sciences, Profsoyuznaya~84/32, 117997 Moscow, Russia \label{inst:RAS} 
    \and Kazan Federal University, Kremlevskaya str.~18, 420008 Kazan, Russia \label{inst:KFU}
    }

%\date{\today}
\date{Received date / Accepted date}

\abstract
{With the progenitors of type Ia supernovae (SNe~Ia) still eluding direct detections, various types of accreting white dwarfs (WDs)  have been proposed as prospective candidates. Among the possibilities, WDs undergoing unstable nuclear burning on their surfaces  have been considered. Although observations and theoretical modelling of Classical Novae generally suggest that  more material is ejected during the explosion than accreted, there is growing evidence that in certain accretion regimes of novae, appreciable mass accumulation by the  WD in the course of unstable nuclear burning may be possible.}
{We propose that statistics of novae in nearby galaxies may be a powerful tool to gauge the role of such systems in producing SNe~Ia.}
{We use multicycle nova evolutionary models to compute the number and temporal distribution of novae that would be produced by a typical SN~Ia progenitor before reaching the Chandrasekhar mass limit ($M_{\rm ch}$) and exploding, assuming that it experienced unstable nuclear burning during its entire accretion history.
We then use the observed nova rate in M31 to constrain the maximal contribution of the nova channel to the SN~Ia rate in this galaxy. 
} 
{The M31 nova rate  measured by the POINT-AGAPE survey is $\approx65$~yr$^{-1}$. Assuming that all these novae will reach $M_{\rm ch}$, we estimate the maximal SN~Ia rate novae may produce, which is $\lesssim 0.1\mbox{--}0.5\times 10^{-3}$~yr$^{-1}$. This constrains the overall  contribution of the nova channel to the  SN~Ia rate at  $\lesssim 2$--$7\%$. 
However, if all  POINT-AGAPE novae do eventually reach $M_{\rm ch}$, one should expect a significant population of fast novae ($t_2\lesssim 10$~days)  originating from the most massive WDs, with the rate of $\sim 200-300$~yr$^{-1}$, which is significantly higher than currently observed. We point out that statistics of such fast novae can provide powerful diagnostics of the contribution of the nova channel to the final stage of mass accumulation by the single degenerate (SD) SN~Ia progenitors.
To explore the prospects of their use, we investigate the efficiency of detecting fast novae as a function of the limiting magnitude and temporal sampling of a nova survey of M31 by a PTF class telescope. We find that a survey with the limiting magnitude  of $m_{R}\approx22$ observing  at least every 2nd night will catch  $\approx 90\%$ of fast novae expected in the SD scenario. Such surveys should be detecting fast novae in M31 at a rate of the order of $\gtrsim 10^{3}\times f$ per year, where $f$ is the fraction of SNe~Ia which accreted in the unstable nuclear burning regime while accumulating the final $\Delta M\approx 0.1~M_\odot$ before the supernova explosion.
}
{}
{}

\titlerunning{Constraining novae as SN~Ia progenitors}
\authorrunning{Soraisam \& Gilfanov}
\maketitle

\section{Introduction}
\label{sec:intro}
The landmark discovery of the accelerating expansion of the Universe came about owing to the use of Type Ia supernovae (SNe~Ia) to measure cosmological distances \citep{Riess, Perlmutter}. With the advent of the era of large-scale surveys yielding large samples of SNe~Ia, the uncertainties in cosmological SN~Ia studies are now dominated not by statistical but systematic effects. The cosmological distance measurements based on SNe~Ia assume that the physical properties of their progenitors remain unchanged with redshift,  although growing evidence points to the contrary  \citep[e.g.,][]{Milne-2015}. Hence, the lack of understanding of the nature of their  progenitors is one of the  important sources of  systematic uncertainties (see \citealt{Howell, Maoz} for a review). 
Although it is established beyond reasonable doubts that these gigantic explosions are a result of the thermonuclear disruption of a carbon-oxygen~(CO) white dwarf~(WD) near the Chandrasekhar mass limit, the details are still being debated. In particular, no consensus has been reached regarding how the WD, whose initial mass is likely to be below $\sim 1~M_\odot$, reaches the Chandrasekhar limit.  To date, there are two major hypotheses -- the single degenerate (SD) scenario, in which a WD gains mass by accreting hydrogen-rich material from a non-degenerate companion before exploding as a SN~Ia (\citealt{Whelan}; see \citet{Wang-2012} for a recent review); and the double degenerate (DD) scenario, in which two CO~WDs coalesce driven by gravitational wave radiation producing a SN~Ia \citep{Webbink,Iben}.

In the SD scenario, the rate at which matter supplied by the donor star  is accreted by the WD  is the critical parameter that determines the fate of this matter.  As the WD in a binary system accumulates material from its companion, nuclear burning of hydrogen is ignited at the base of the accreted hydrogen-rich envelope after the critical temperature and pressure are reached. Calculations of several groups have led to the conclusion  that stable nuclear burning of the accreted material at the same rate as it is supplied by the donor star is possible only in a rather narrow range of the mass accretion rates around $\sim{\rm few}\times10^{-7}$~M$_\odot$~yr$^{-1}$ \citep{Nomoto, Wolf, Kato-2014}. It has been argued that this is the regime allowing the most efficient build-up of mass of the WD. Due to the energy released in the hydrogen fusion, the WD becomes a powerful source of soft X-ray emission, the so-called supersoft X-ray source (SSS; \citealt{Heuvel}; \citealt{Kahabka}). At larger mass accretion rates, not all the matter can be processed in the nuclear fusion and it has been argued, whether the accreted envelope expands dramatically leading to a red-giant-like configuration \citep{Cassisi} or a radiation driven wind blows away the excess mass (wind-regime; \citealt{Hachisu-1996}). At  lower mass accretion rates, below the stability strip,  conditions at the base of the  envelope of accreted material are insufficient for steady hydrogen fusion and it undergoes  regular thermonuclear runaways, resulting in nova explosions. The explosions may be accompanied by significant mass loss from the system (e.g., \citealt{Prialnik}).

While numerous attempts to find the progenitor of individual SN~Ia have not (yet) yielded convincing detections, an alternative avenue has been recently explored, aimed to constrain the {\em overall populations} of accreting WDs in galaxies. In particular,  \citet{Gilfanov-2010} demonstrated  that the observed soft X-ray luminosity of  early-type galaxies is too low to allow a significant population of hot ($T\gtrsim5\times10^5$~K) accreting WDs with stable nuclear burning, required to explain the SN~Ia rates in these galaxies. Populations of cooler WDs\footnote{Such cooler sources may appear in the case of lower mass WDs or a considerable expansion of the WD photosphere, cf. the wind-regime.} have been constrained  by \citet{Woods-a, Woods-b} and \citet{Johansson}  using the strength of  He II recombination lines in the line emission spectra of passively evolving galaxies. This work effectively excluded the parts of the parameter space corresponding to accreting WDs in and above the stability strip, at least in passively evolving galaxies.

Below the stability strip, theoretical modelling of nova explosions has demonstrated that at mass accretion rates $\lesssim 10^{-8}-10^{-7}$~M$_\odot$~yr$^{-1}$, the entire accreted mass is likely to be lost from the system during the explosion  (e.g., \citealt{Prialnik, Yaron}). However, close to the stability line, $\dot{M}\sim 10^{-7}$~M$_\odot$~yr$^{-1}$, nova explosions are relatively weak and are not accompanied by significant mass-loss; therefore, mass accumulation by the WD may be possible in this regime \citep{Hillman-2015}.  
Moreover,  work by  \citet{Starrfield-2014}, \citet{Hillman-2015} and others has contested the entire existence of the  stability strip. Instead, they suggested that  nuclear burning is unstable in the entire mass accretion rate range upto very high rates, where  common envelopes form. In this picture, the nuclear burning  proceeds in the form of flashes caused by regular thermonuclear runaways, with significant net mass accumulation by the WD.  These results inspired a corresponding class of SN~Ia progenitor models involving novae of various types 
(e.g., \citealt{Starrfield-1985, Hachisu-2001, Hillman-2015}). However, \citet{Gilfanov-2011} pointed out that presence of a significant population of such systems with unstable nuclear burning in galaxies will result in nova rates by far exceeding the observed values.

In this paper, we investigate the  proposition  made by \citet{Gilfanov-2011} in depth. In Sec.~\ref{sec:rate},  we compute the number of novae expected to be produced by a typical SD progenitor accreting in the unstable nuclear burning regime. To this end, we use the multicycle nova models of \citet{Prialnik}, \citet{Yaron} and \citet{Hillman-2015}. Using the predicted number of novae per SN~Ia and the nova rate in M31 measured by the POINT-AGAPE survey \citep{Darnley}, we estimate the maximal contribution that novae can make to the SN~Ia rate in Sec.~\ref{sec:nov_M31}. We then point out that, should the nova channel be responsible for a significant fraction of SNe~Ia, the vast majority of novae would have short decay times; thus, an even more sensitive diagnostic should be provided by observations of fast novae. We  address the completeness of nova surveys for fast novae in Sec.~\ref{incompleteness} and formulate requirements for future high cadence surveys aimed to further examine the role of novae as SN~Ia progenitors. Our results are discussed in the broader context of SN~Ia models in Sec.~\ref{disc2}. In this section, we also  address  the dependence of our results  on the uncertainties of the multicycle nova models; in particular, we compare \citet{Yaron} models used in this paper with the calculations of other groups and with observations.  We conclude in Sec.~\ref{conclusion}.

\section{Relation between supernova and nova rates}\label{sec:rate}

Assuming that an accreting WD spends some fraction of its accretion history  in the unstable nuclear burning  regime,  one can  estimate the number of novae $n_{\rm nov}$ it will produce while increasing its mass by $\Delta{M}_{\rm wd}$ from the following obvious relation
\begin{equation}
\label{eq:crude_rate}
\Delta{M}_{\rm acc} n_{\rm nov}\sim\Delta{M}_{\rm wd},
\end{equation}
where $\Delta{M}_{\rm acc}$ is the net mass gain by the WD per one nova explosion cycle.

However, the  ignition mass of the nova, and hence  $\Delta{M}_{\rm acc}$, depend on  parameters of the WD and the binary system.  To the first approximation,  the main parameters are the WD mass $M_{\rm WD}$, its temperature $T_{\rm WD}$, and the accretion rate $\dot{M}$ \citep{Yaron}, and  we can rewrite Eq.~\eqref{eq:crude_rate} more accurately as
\begin{equation}
\label{eq:prec_rate}
n_{\rm nov}=\int\limits_{M_{\rm i}}^{M_{\rm ch}}\frac{dM_{\rm WD}}{\Delta{M}_{\rm acc}(M_{\rm WD},\dot{M},T_{\rm WD})}.
\end{equation} 
Eq.~\eqref{eq:prec_rate} gives the number of novae produced by an SD SN~Ia progenitor while increasing its mass from some initial value $M_{\rm i}$ to  the Chandrasekhar mass $M_{\rm ch}$, assuming that the nuclear burning on the WD surface proceeds in the unstable regime.

The majority of mutlicycle nova evolution models give ignition mass rather than  the accreted mass, as the former is much more straightforward to compute \citep[e.g.,][]{Townsley}. For this reason we reformulate Eq.~\eqref{eq:prec_rate} in terms of the ignition mass $\Delta{M}_{\rm ign}$. Because of the mass loss  during the nova explosion, $\Delta{M}_{\rm acc}\le \Delta{M}_{\rm ign}$, and therefore we obtain a lower limit on the number of novae produced by one single degenerate SN~Ia progenitor
\begin{equation}
\label{total_rate}
n_{\rm nov}\geq\int\limits_{M_{\rm i}}^{M_{\rm ch}}\frac{dM_{\rm WD}}{\Delta{M}_{\rm ign}(M_{\rm WD},\dot{M},T_{\rm WD})}.
\end{equation}

\begin{figure}
\includegraphics{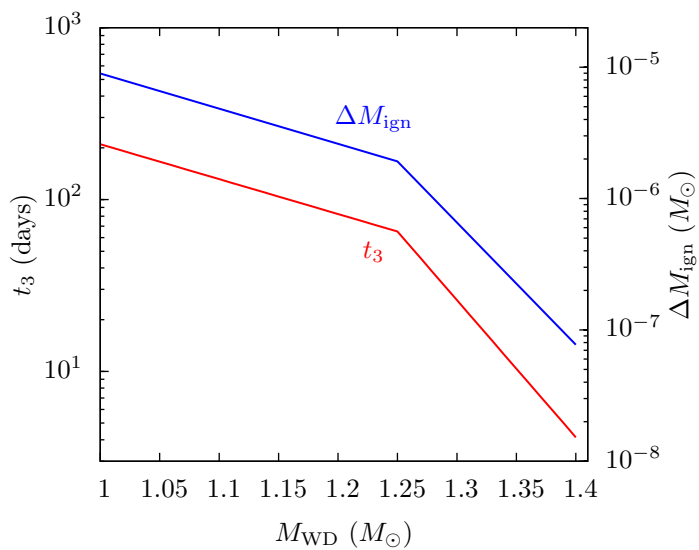}
\caption{Variation of the ignition mass of the nova $\Delta{M}_{\rm ign}$ and the mass-loss timescale $t_{ml}$ ($\approx$ decline time of the outburst by 3 magnitudes from optical peak, $t_3$) with the mass of the WD for the accretion rate $\dot{M}=10^{-7}~M_{\odot}{\rm yr}^{-1}$ and WD core temperature $10^{7}$~K, from \citet{Yaron}.}\label{fig:ignition}
\end{figure}

Eq.~\eqref{total_rate} gives an estimate of the total number of novae regardless of their individual characteristics. On the other hand, we can see from Fig.~\ref{fig:ignition} that as the WD mass increases towards the Chandrasekhar limit, $\Delta{M}_{\rm ign}$ decreases as well as $t_{3}$ -- decline time of the optical light curve from the peak by 3~magnitudes. Therefore, as the WD mass  grows, it produces more frequent nova explosions with shorter decline times, i.e., the majority of the novae relevant to the SN~Ia progenitor problem should be characterised by fast decline. Thus, a more detailed diagnostic can be provided by the distribution of novae over the decline time of their light curves.

To obtain the latter, we note that in the multicycle  nova evolution models of \citet{Yaron}, for fixed mass accretion rate and WD core temperature, the light curve decay timescale 
is determined solely by the WD mass. Therefore,  the cumulative distribution of novae over decay time is given by
\begin{equation}
\label{eq:cum_dist}
n_{\rm nov}\left(\leq t_n \right) \, \ge \int\limits_{M(t_n)}^{M_{\rm ch}}\frac{dM_{\rm WD}}{\Delta{M}_{\rm ign}(M_{\rm WD},\dot{M},T_{\rm WD})}.
\end{equation}
Here $t_{n}$ is the time to decline by $n$~mag from peak ($n=2,3$), $n_{\rm nov}(\leq t_{n})$ is the cumulative number of novae with decline time less than or equal to $t_{n}$ and $M(t_{n})$ is the WD mass corresponding to the given decline time $t_{n}$ (for the given values of $\dot{M}$ and $T_{WD}$). As before, the inequality sign in Eq.~\eqref{eq:cum_dist} reflects the fact that the net accreted mass can be smaller than the envelope ignition mass. The corresponding differential distribution is given by
\begin{equation}
\label{eq:diff_rate}
\frac{dn_{\rm nov}(t_{n})}{dt_{n}}\, \ge \,  \frac{1}{\Delta{M}_{\rm ign}(M_{\rm WD},\dot{M},T_{\rm WD})}\frac{dM(t_{n})}{dt_{n}}.
\end{equation}
As before, this distribution gives the number of novae with particular temporal properties per one type Ia supernova.

In the following,  we will use the results of the multicycle nova evolutionary calculations by \citet{Yaron}. Their results for the envelope ignition mass and $t_3$ timescale of the light curve are shown in Fig.~\ref{fig:ignition}.
We carry out calculations for three values of the mass accretion rates -- $10^{-7}$, $10^{-8}$ and $10^{-9}~M_{\odot}~{\rm yr}^{-1}$ and assume WD core temperature of $10^7$~K. Our choice is explained below.  
In addition, some of the calculations in Sec.~\ref{sec:nov_M31} are done for the mass accretion rate of $5\times10^{-7}~M_{\odot}~{\rm yr}^{-1}$, which is not tabulated in \citet{Yaron}. To  this end,  we use the results of \citet{Hillman-2015} that are based on a modified version of the code of \citet{Prialnik}.
We  estimate the nova ignition masses  using their plot of the nova cycle duration against the mass accretion rate (Fig.~2 in \citealt{Hillman-2015}). As a proxy for the nova decline time $t_3$ we use the flash duration plotted in their Fig.~7. We verified that the  so obtained nova ignition masses and $t_3$ timescales are consistent (albeit not identical) with the interpolation of \citet{Yaron} results.

\begin{figure*}
\includegraphics{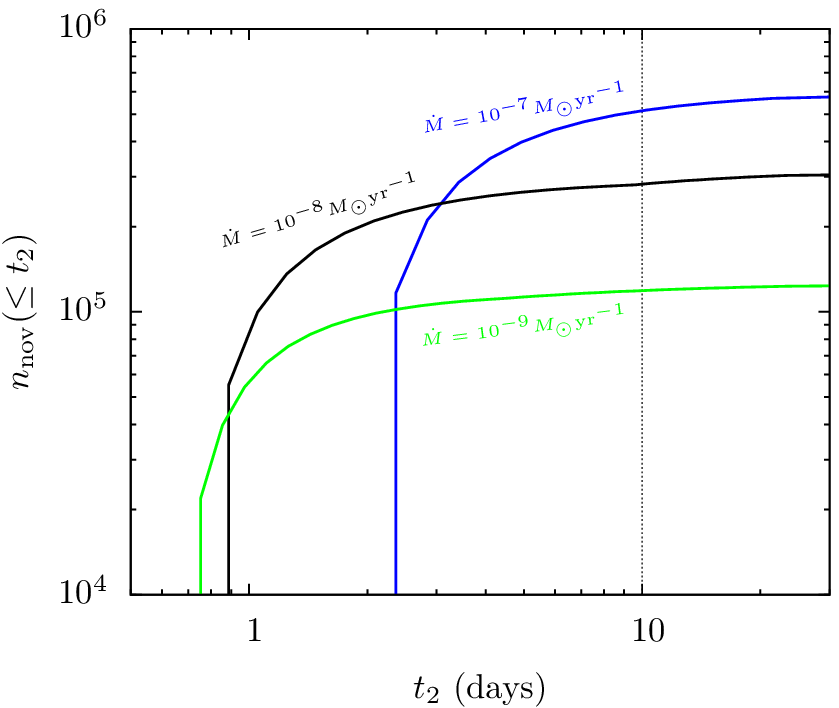} \hfill \includegraphics{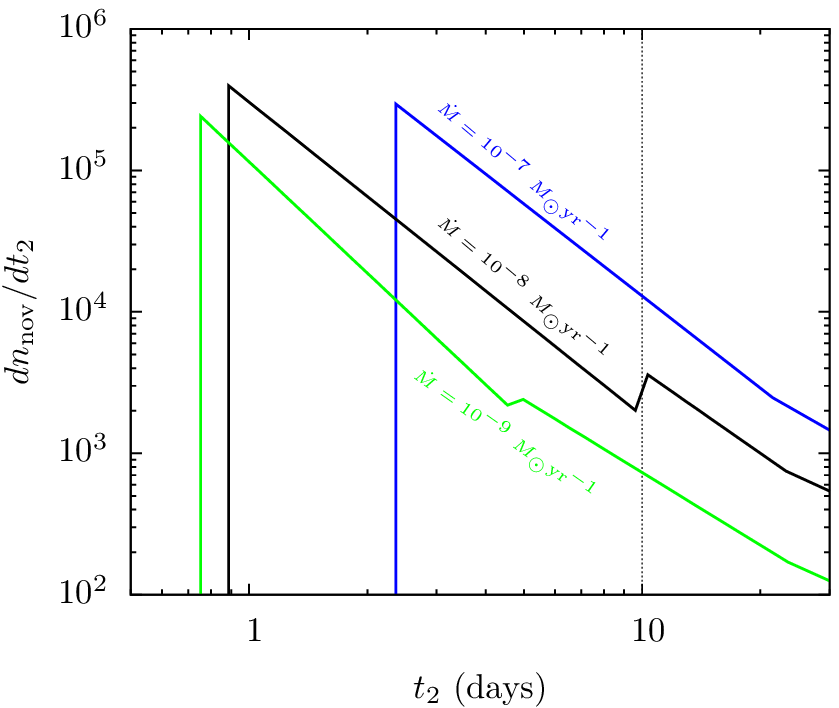}
\caption{Cumulative ({\it left}) and differential ({\it right}) $t_2$ distributions of novae  produced by one successful SN~Ia progenitor, assuming that it accreted in the unstable nuclear burning regime throughout its entire accretion history. The numbers near the curves indicate the mass accretion rate.  The discontinuities  in the differential distributions  are caused by the discontinuity in the slope of log-linear interpolation of $t_2$ and $M_{\rm WD}$ values from \citet{Yaron} tabulation. They do not affect the cumulative distribution  in any significant way, as can be seen in the left panel. The sharp drop in the distributions at small $t_2$ values corresponds to the fastest novae produced by the WD near the Chandrasekhar mass limit.}\label{fig:cum_diff}
\end{figure*}

The mass loss during the nova explosion becomes more significant at lower mass accretion rates \citep[e.g.,][]{Yaron, Hillman-2015}. Therefore, in the context of the problem of SN~Ia progenitors, only relatively high mass accretion rates ($\sim 10^{-7}~M_{\odot}~{\rm yr}^{-1}$) are relevant. We will therefore assume the mass accretion rate of $10^{-7}$~$M_{\odot}~{\rm yr}^{-1}$ in our baseline configuration, but also consider smaller values of $10^{-8}$ and $10^{-9}~M_{\odot}~{\rm yr}^{-1}$ in order to investigate the trends with $\dot{M}$. Some of the calculations in Sec.~\ref{sec:nov_M31} are done for $\dot{M}=5\times10^{-7}~M_{\odot}~{\rm yr}^{-1}$ to allow direct comparison of predictions of the \citet{Hillman-2015} model with observations.

Our choice of the WD core temperature is motivated  by the results of \citet{Townsley}, who studied the effect of accretion on the thermal state of the WD and found the equilibrium WD core temperature of $\sim8$--$10\times10^{6}$~K for accretion rates in the range $10^{-9} -10^{-8}~M_{\odot}~{\rm yr}^{-1}$, the  temperature increasing with $\dot{M}$. Their calculations covered  only  the mass accretion rates typical for classical novae and did not extend beyond $10^{-8}~M_{\odot}~{\rm yr}^{-1}$. However, at high mass accretion rates, $\gtrsim 10^{-7}$~$M_{\odot}~{\rm yr}^{-1}$, the properties of nova outbursts do not strongly  depend on the core temperature because the overlying hot He layer from previous outbursts acts as a heat barrier \citep{Townsley, Wolf}. We therefore assume the core temperature of $10^7$~K in our baseline configuration. We further investigate dependence of our results on the core temperature in Sec.~\ref{sec:nov_M31}.

\citet{Yaron} provide two timescales characterizing the light curve decay rate -- the duration of the mass-loss phase $t_{\rm ml}$ and the decline time of the bolometric luminosity by 3 mag from maximum $t_{3,{\rm bol}}$. From an observational point of view, however, the timescale of interest is the decline time of the optical light from the nova. From the comparison of the results from \citet{Yaron} with observations, it is known that $t_{\rm ml}$ and $t_{3,{\rm bol}}$ bracket $t_{3}$ \citep{Prialnik, Yaron, Kasliwal}. The shorter of these two, $t_{\rm ml}$, is much closer to the observed $t_{3}$ than $t_{3,{\rm bol}}$. We will, therefore, proceed with $t_{\rm ml}$ as an approximation to $t_{3}$. We will investigate how our results change  if we use the longer $t_{3,{\rm bol}}$ in Sec.~\ref{sec:nov_M31}.  In practice,  for extragalactic novae, where the survey sensitivity\footnote{It is generally limited by the unresolved surface brightness of the host galaxy rather than the limiting magnitude of the survey as such (see Section \ref{incompleteness}).} becomes an issue, it is easier to measure $t_{2}$ (the time to decline by 2~mag from peak) than $t_{3}$.  For fast novae, the two quantities are approximately related via $t_{2}\approx t_{3}/2.1$, while for the slow ones via $t_{2}\approx t_{3}/1.75$, with the transition at $t_{3}=50$~days following Duerbeck (in \citet{Bode}). We will mostly use $t_2$ throughout the rest of the paper.

For each value of the mass accretion rate we log-linearly interpolate the ignition mass $\Delta{M}_{\rm ign}$, $t_{\rm ml}$ and $t_{3, {\rm bol}}$ between the grid values of the WD mass (0.4, 0.65, 1.0, 1.25 and 1.4~$M_{\odot}$)  in \citet{Yaron}. The lower integration limit in Eq.~\eqref{eq:cum_dist} is conservatively assumed to be equal to $1.0~M_{\odot}$. For $\dot{M}=10^{-7}$~M$_\odot$~yr$^{-1}$, this corresponds to slow novae with $t_3\approx 200$~days. As the differential distribution of novae over $t_2$ is sufficiently steep, $dn_{\rm nov}/dt_2\propto t_2^{-2}$ (see discussion of Fig.~\ref{fig:cum_diff} below), the choice of the initial WD mass is not very important when considering the integrated nova rates, as long as it is not too close to the Chandrasekhar mass.

The so computed cumulative and differential $t_2$ distributions of  novae  are shown in Fig.~\ref{fig:cum_diff}. As expected, a significant fraction of novae produced by a successful SD SN~Ia progenitor have short $t_2$ times.  For example, in our baseline calculation ($\dot{M}=10^{-7}~M_\odot$~yr$^{-1}$) we obtain the total predicted number of novae per one SN~Ia $\sim 6.5\times 10^5$. Of these, more than $60\%$ have $t_{2}\lesssim5$~days and about $80\%$ have $t_2$ shorter than 10~days. The predicted number of novae grows with the mass accretion rate, reaching $\sim 7.7\times 10^5$ for $\dot{M}=5\times 10^{-7}$~M$_\odot$~yr$^{-1}$.

\section{Statistics of novae in M31}\label{sec:nov_M31} 

M31 has been a hot spot for the observation of novae since the work of \citet{Hubble}. With ample detections, we have chosen this galaxy  for our analysis. We start this section by reviewing recent supernova and nova rate measurements in M31, and then proceed with constraining the contribution of the nova channel to the observed SN~Ia rate using the results of the previous section.

\subsection{SN~Ia rate in M31}\label{snia_in_M31}

The morphological type of M31 is Sb \citep{Vaucouleurs-1991}, for which \citet{Mannucci} give the stellar mass specific SN~Ia rate of $6.5\times10^{-4}$~SNe~${\rm yr}^{-1}$ per $10^{10}~M_\odot$. However, Hubble morphological classes are rather broad --  indeed, for the adjacent morphological type Sbc/d, the quoted SN~Ia rate is higher by a factor of $\sim 2.5$.  Therefore, there must inevitably be some spread in the SN~Ia rates between galaxies of the same morphological type.  In the \citet{Mannucci} work, the morphological type of the galaxy is used as a proxy for its star formation rate (SFR). Indeed, to the first approximation, in the absence of detailed knowledge of the star formation history of the galaxy, the mass specific SN~Ia rate is determined by the current SFR (e.g., \citealt{Sullivan-2006}; see also \citealt{Maoz}). A more accurate and continuous characterisation of the current SFR of the galaxy is its colour, albeit also with considerable spread. In particular, \citet{Mannucci} used $B-K$ colour to quantify dependence of the supernova rate on SFR. The extinction corrected $B-K$ colour of M31 within $\sim 10\mbox{--}20$ kpc from the galactic center is, $B-K\approx 3.5~{\rm mag}$ \citep{Battaner-1986}. From Fig.~5 in \citet{Mannucci} we find the mass specific SN~Ia rate of $\sim 1.0\times10^{-3}$~SNe~${\rm yr}^{-1}$ per $10^{10}~M_\odot$, i.e., somewhat higher than inferred from its morphological type. This number is compatible with the more recent result of \citet{Li-2011}, who obtained the mass specific SN~Ia rate of  $(0.85^{+0.13}_{-0.12})\times10^{-3}$~SNe~${\rm yr}^{-1}$ per $10^{10}~M_\odot$ in their $B-K=3.4$--$3.7~{\rm mag}$ bin for a galaxy of stellar mass $1.1 \times10^{11}~M_\odot$ (see below). Of course, one can also use the current SFR value directly. The SFR estimates for M31 are in the range $\approx 0.4\mbox{--}0.8~M_{\odot}~{\rm yr}^{-1}$ \citep{Barmby, Devereux-1994}, and with the \citet{Sullivan-2006} calibration, we obtain the SN~Ia rate of $(0.5\mbox{--}0.6)\times10^{-3}$~SNe~${\rm yr}^{-1}$ per $10^{10}~M_\odot$, also compatible with the above numbers.

From the physical point of view, the SN~Ia rate of the galaxy is determined by the convolution of its star formation history with the delay time distribution (DTD) of SNe~Ia \citep{Maoz-2012}. As the former is poorly known, we will use the DTD value at the delay time equal to the mean stellar age of M31 to  estimate its SN~Ia rate.
\citet{Olsen-2006} found both the bulge and inner disk of M31 to be dominated by old (6--10~Gyr) stellar population, and \citet{Brown-2006} found the outer disk to be dominated by 4--8~Gyr old stars. Then, taking the mean age for the M31 stellar population to be 8~Gyr, and using the delay time distribution of \citet{Totani-2008}  we obtain the mass specific rate of $\approx 0.7\times10^{-3}$~SNe~${\rm yr}^{-1}$ per $10^{10}~M_\odot$.

Thus, different estimations give approximately consistent values of the  specific SN~Ia rate in M31, in the range of  $\approx (0.5\mbox{--}1.0)\times10^{-3}$~SNe~${\rm yr}^{-1}$ per $10^{10}~M_\odot$. In the following, we will conservatively use the rate based on the \citet{Mannucci} result for Sb galaxies, i.e., $0.65\times10^{-3}$~SNe~${\rm yr}^{-1}$ per $10^{10}~M_\odot$, which is one of the lower rate estimates from above. As the nova rates are directly proportional to the SN~Ia rate (e.g., Eq.~\eqref{total_rate}), any higher SN~Ia rate will only make our conclusions stronger. With the stellar mass of M31 of $1.1\times10^{11}~M_\odot$ \citep{Barmby-2007} we obtain its SN Ia rate, $\dot{N}_{\rm SNIa}=7.15\times10^{-3}~{\rm yr}^{-1}$, which we will use in our calculations below.

\subsection{Nova rate in M31}
\label{sec:constraints}

Altogether, there are more than 900 \footnote{\label{MPE_table}\url{http://www.mpe.mpg.de/~m31novae/opt/m31/M31_table.html} \citep{Pietsch-2007}} novae detected in the direction of M31 (e.g., \citealt{Hubble, Arp}; see also \citealt{Capaccioli}). However, the majority of the existing nova catalogs lack accurate incompleteness analysis, that renders them unsuitable for use in our calculations. 

Completeness of a nova survey is determined by the usual factors, such as spatial variation of the sensitivity caused by incomplete coverage of the survey, variation of the surface brightness of the galaxy, extinction, etc.  In addition, it is determined by the factors related to the transient nature of the sought objects, such as temporal sampling of the survey and variation in the light curve morphology. The most thorough completeness analysis to date among the nova surveys was performed for the nova catalog produced in the course of  the recent POINT-AGAPE (Pixel-lensing Observations with the Isaac Newton Telescope -- Andromeda Galaxy Amplified Pixels Experiment) survey  \citep{Darnley-2004}. The POINT-AGAPE nova catalog was produced by an automated detection pipeline, thus permitting an objective characterisation of its completeness (many, even relatively recent nova surveys relied on some form of a visual inspection of the images and/or light curves, which makes their completeness difficult to  compute accurately).  \citet{Darnley} carried out a careful analysis of the completeness of the detection pipeline and the survey itself. To this end, they seeded the raw POINT-AGAPE data with resampled light curves of their detected novae. This allowed them to compute the completeness of the POINT-AGAPE nova catalog and to obtain a robust estimate of the underlying global nova rate in M31. As a result of these analyses they produced the global nova rate of $65^{+16}_{-15}$~yr$^{-1}$.

\subsection{Contribution of novae to the SN~Ia rate}\label{sec:fast_novae}

With the nova rate known, one can now estimate the maximal SN~Ia rate these novae can produce as follows
\begin{equation}
\label{eq:max_sn_rate}
\dot{N}_{\rm Ia, nov}\le \frac{\dot{N}_{\rm nov}}{n_{\rm nov}},
\end{equation}
where $\dot{N}_{\rm Ia, nov}$ is the SN~Ia rate that may be produced by novae, $n_{\rm nov}$ is the number of novae produced by one successful SN~Ia progenitor (cf.~Eq.~\eqref{total_rate}) and $\dot{N}_{\rm nov}$ is the observed nova rate. Note that Eq.~\eqref{eq:max_sn_rate} gives an upper limit on the nova contribution to the SN~Ia rate for at least two reasons: (i)~Eq.~\eqref{total_rate} gives only the lower limit of the number of novae per SN~Ia, as discussed in Sec.~\ref{sec:rate}; (ii)~obviously, not all novae reach the Chandrasekhar mass limit.

However, before the nova rate of $65~{\rm yr}^{-1}$ can be plugged into Eq.~\eqref{eq:max_sn_rate}, the following should be considered. The POINT-AGAPE nova catalog does not contain very fast novae with $t_{2}\lesssim10$~days in the $r'$ band. Therefore, as discussed in \citet{Darnley}, their completeness modelling is only sensitive to novae with an $r'$-band  $t_2$ between  9.80 days (the fastest nova in their catalog) and  213.12 days  (the slowest nova). In this $t_2$ range, there was no strong evidence suggesting large variation  in the completeness as a function of $t_2$,  the only variation being, as expected, spatial (Matt~J.~Darnley, private communication). Therefore, for a fair comparison, $n_{\rm nov}$ in Eq.~\eqref{eq:max_sn_rate} should be the number of novae with light curve decay times in the range $10\la t_2\la 213$~days. Due to the steepness of the  differential distribution $dn_{\rm nov}/dt_2$ (Fig.\ref{fig:cum_diff}), the latter is nearly equivalent to $t_2\ge 10$~days.\footnote{Note that since the red-bands contain the H-alpha emission line, which declines more slowly than the continuum, the quoted $t_2$ time may be somewhat overestimated compared to the $t_{2}$ measured  in the V-band (see \citealt{Darnley}). An accurate account for this effect is beyond the scope of this paper; we note, however, that it will increase the predicted rates (cf. Fig.~\ref{fig:cum_diff}) and  will result  in an even tighter constraint.}

\begin{figure}
\includegraphics{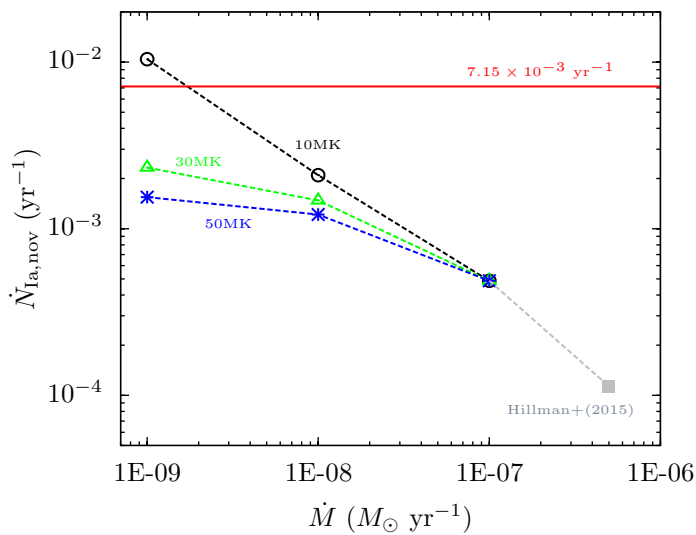}
\caption{Maximal contribution of novae to the SN Ia rate of M31 as a function of assumed mass accretion rate for different values of  WD core temperature as indicated by the numbers near the curves. The  point  at $\dot{M}=5\times10^{-7}$ M$_\odot$~yr$^{-1}$ (shown in grey) is computed from the models of \citet{Hillman-2015}; other points are computed based on \citet{Yaron} nova models. The red solid line shows the M31 supernova rate.}\label{fig:acc_mdot}
\end{figure}

The maximal supernova rate computed from  Eq.~\eqref{eq:max_sn_rate} is shown in Fig.~\ref{fig:acc_mdot}. As one can  see, our baseline model ($\dot{M}=10^{-7}~M_{\odot}~{\rm yr}^{-1}$) predicts the maximal SN~Ia rate of  $\approx 5.0\times 10^{-4}~{\rm yr}^{-1}$. For the mass accretion rate of $\dot{M}=5\times 10^{-7}~M_{\odot}~{\rm yr}^{-1}$ considered by \citet{Hillman-2015}, the maximal SN~Ia rate becomes $\approx 1.1\times 10^{-4}~{\rm yr}^{-1}$.
Interestingly, for lower mass accretion rates, observed population of novae could in principle explain a larger fraction of SNe~Ia, and  for very low rates, $\dot{M}\la 10^{-9}~M_{\odot}~{\rm yr}^{-1}$, and low WD temperature, the predicted rate is compatible with the observed value. However, such low mass accretion rates are  believed to be irrelevant as regards the nature of SN~Ia progenitors  (Sec.~\ref{sec:rate}). For the mass accretion rates in the range $\sim (1-5)\times 10^{-7} ~M_{\odot}~{\rm yr}^{-1}$, typically considered in this context, the maximal contribution of novae to the observed SN~Ia rate is limited to $\approx 2-7\%$.

The upper limit on the  contribution of novae to the SN~Ia rate decreases by a few times if we use $t_{3,{\rm bol}}$ instead of $t_{\rm ml}$  in Yaron et al. models. In this case, the  supernova rate predicted in our baseline model is $\approx 1.8\times 10^{-4}~{\rm yr}^{-1}$, i.e., $\sim 40$ times smaller than the observed SN~Ia rate. At lower mass accretion rates and WD temperatures the maximal supernova rate is $\sim $5--10 times short of the observed value.

In the above calculation, we ignored the possible difference of the observed novae in the WD composition. In particular, some fraction of the observed nova outbursts may be hosted by oxygen-neon (ONe) WDs (\citealt{Truran-1986, Ritter-1991, Gil-Pons-2003, Shore-ONe}). For example, for the Galactic novae, \citet{Gil-Pons-2003} estimated this fraction to be  about $\sim 1/3$ of nova outbursts.  The ONe WDs  are known to undergo accretion-induced collapse upon reaching the Chandrasekhar mass limit, rather than producing SNe~Ia and, therefore,  should be excluded from our calculation of the contribution of novae to the supernova rate.  
This is not possible, as the composition of the WD host in the majority of observed novae is unknown. However, Eq.~\eqref{eq:max_sn_rate} gives an upper limit for the nova contribution to the  supernova rate and inclusion of some number of ONe novae in the nova rate does not invalidate it. Although the upper limit may be tightened somewhat by using the (unknown) pure CO nova rate, the \citet{Gil-Pons-2003} results suggest that the improvement will not be dramatic. It should also be noted that the POINT-AGAPE sample is dominated by the relatively slow novae with $t_2\ga 10$ days, which, according to Yaron et al. models, are hosted by relatively lower mass WDs (cf.~Fig.~\ref{fig:ignition}). Therefore the POINT-AGAPE sample should be less contaminated by ONe novae (which are typically hosted by  more massive WDs) than the  overall population of novae.

\subsection{Fast novae}
\label{sec:new}

About $\sim 80\%$ of the novae produced by a typical successful (unstably burning) SD SN~Ia progenitor have short decay times, $t_2\la 10$ days  (Sec.~\ref{sec:rate}, Fig.~\ref{fig:cum_diff}). If a fraction $f$ of the total number of SN~Ia progenitors in M31 accrete in the unstable nuclear burning regime while accumulating their final $\Delta M\approx 0.1 M_\odot$  (i.e., from $M\approx 1.3~M_\odot$ to $M_{\rm ch}$),  fast novae with $t_2\la 10$ days\footnote{Recall that in \citet{Yaron} models, the $t_2\approx10$~days  corresponds to the WD mass of $\approx 1.3$~M$_\odot$, with the more massive WDs producing faster novae.} should be produced at the rate of $\ga 3.6\cdot 10^3\times f$, assuming $\dot{M}=10^{-7}$~M$_\odot$/yr. For example, should all the novae from the POINT-AGAPE sample reach the Chandrasekhar mass limit, the fast novae would be produced at the rate of $\sim 200-300$ yr$^{-1}$.

Such fast novae have not been detected in the POINT-AGAPE survey, either because they are rare in M31 or the survey was not sensitive to them, or a combination of these two reasons \citep{Darnley}. From results of other surveys, we  know that some number of  such fast novae do exist, an example being the famous M31N~2008-12a \citep{Shafter-2012, Darnley-2014, Henze-2014, Tang} (see also the MPE optical nova catalog from footnote~\ref{MPE_table}),  however their true frequencies in  the bulge and the disk of the galaxy remain to be determined.

From the above it is obvious  that statistics of fast novae could provide a powerful tool to investigate the populations of  massive  WDs with unstable nuclear burning and to further constrain their contribution to the observed SN~Ia rates.  Accurate determination of their frequency  is, however,  hindered by the difficulty of their detection due to their short lifetimes, demanding surveys of very high cadence. For example, the famous M31 nova, M31N~2008-12a was discovered as a recurrent nova only in 2008, despite the fact that it explodes every year. This is discussed in the following section, where we investigate how completeness of a nova survey depends on its temporal sampling.

\begin{figure*}
\includegraphics{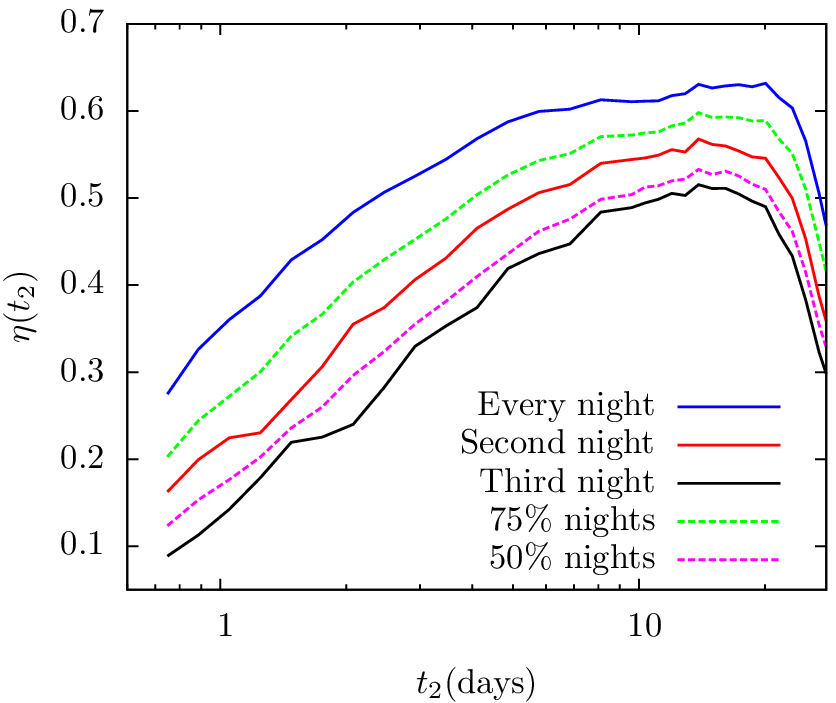} \hfill \includegraphics{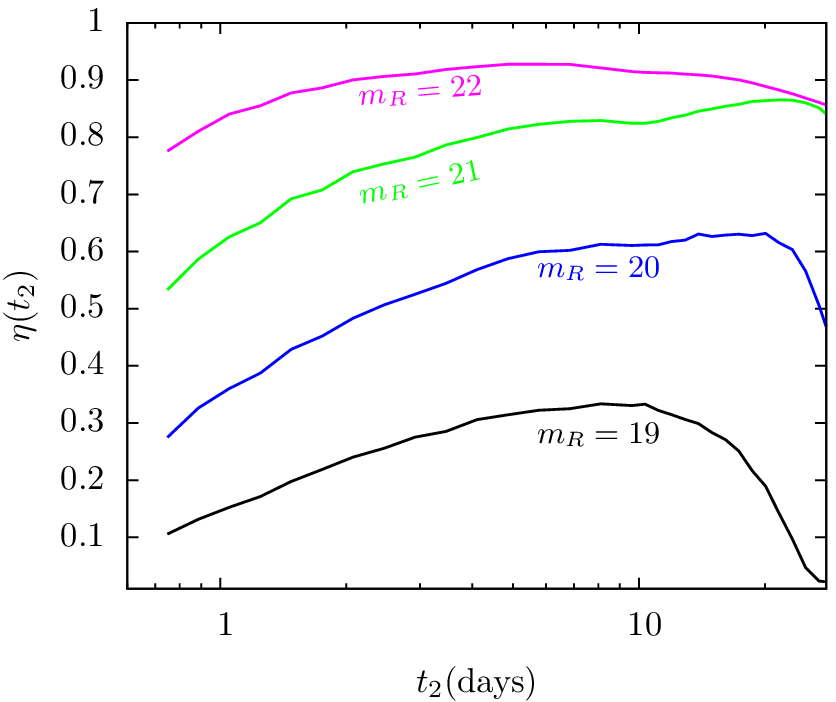}
\caption{Detection efficiency for novae in M31 as a function of their decline time $t_{2}$ for  surveys  with the limiting magnitude of $m_{R}=20$ and different temporal sampling ($left$) and for surveys of different limiting magnitudes with observations performed every night ($right$).  The decline of the detection efficiency for slow novae is caused by their lower peak magnitudes and the finite time span of the simulated survey (see Sec.~\ref{incompleteness}).
}\label{fig:mock}
\end{figure*}

\section{Temporal sampling and completeness of nova surveys}\label{incompleteness} 

In this section we investigate how efficiently fast novae can be detected in surveys of various sensitivity and cadence.   Our  goal is to identify  typical requirements with respect to the temporal sampling and limiting magnitude of a modern CCD survey aimed to characterise the  population of fast novae, rather than to substitute the actual completeness calculations. The latter  should  be performed taking into account the characteristics of the particular survey and parameters of its detection pipeline. We therefore do not include in our calculations the full complexity of the light curve shapes, replacing it with a simple template (albeit derived from observed nova light curves), with its peak magnitude drawn from the range sampled by observed novae. We do, however, take into account spatially varying  internal extinction in M31 and contribution of its unresolved surface brightness to the statistical noise in the image, i.e., to the sensitivity in detecting novae. We conduct our simulations for a PTF-class telescope (1.2 m Samuel Oschin Telescope at the Palomar Observatory).

We  employ the following procedure. Firstly, based on observed novae, we produce a scalable light curve template (Appendix \ref{template}), which we use to model the light curve of a nova with a given peak magnitude and $t_2$ time.  In order to draw the peak magnitude of the nova with the given $t_2$, we produce an analog of the classic MMRD relation  from a large, albeit heterogenous, set of observational data, with the main goal to sample the range of observed magnitudes as fully as possible  (Appendix \ref{mmrd}).  
We then perform Monte-Carlo simulations of the nova detection process. 
In these simulations, for each value of $t_2$ we randomly seed a large number of novae distributed across the face of the galaxy and determine how often they are detected in a survey of a given limiting magnitude and temporal sampling.\footnote{In some respect our approach is similar to the one used by \citet{Darnley} for their completeness analysis, with the difference that they used the very same novae detected in their survey, whereas we are using average statistical properties of a large compilation of novae.} The simulations are carried out in {\it R}-band as it is the band used in many recent nova surveys of M31.

To determine the nova detection sensitivity, we consider the following.
The noise in an image pixel containing $S=S_{\rm M31}+S_{\rm sky}$ counts (in DN unit) accumulated during the exposure time $t_{\rm exp}$ can be expressed as
\begin{equation}
\sigma=\frac{\sqrt{(S_{\rm M31}+S_{\rm sky})\times g+\sigma_{\rm RDN}^{2}+d\times t_{\rm exp}}}{g},
\label{eq:ptf}
\end{equation}
where $g$ is the gain, $\sigma_{\rm RDN}$, the readout noise and $d$, the dark current \citep[see][]{Steve-2006}. For these, we assume typical parameters of the PTF survey \citep{Law}-- 1.6~e-/DN, 12~e- and 0.1~e-/sec, respectively. For the $S_{\rm sky}$ we assumed typical Palomar sky brightness for photometric grey nights \citep{Law, Laher-2014}. For the  $S_{\rm M31}$, we use the SDSS mosaic image of M31 from \citet{Tempel-2011}, which we convert from Sloan-{\em r} to {\em R} band using the relation from \citet{Blanton-2007}. We run this image through SEXTRACTOR \citep{Bertin-1996} to generate the unresolved surface brightness map of the galaxy.  Since the SDSS mosaic image is sampled at a larger pixel scale ($3.96''$) than that of the PTF ($1.01''$), we reduce the pixel counts $S$ from the SDSS image by $\approx 15.4$. 
The radius of the aperture used for measurement of a star is typically of the order of its FWHM \citep[e.g.,][]{Mighell-1999}, which is $\approx 2''$ for PTF images \citep{Law}.  We therefore multiply the pixel $\sigma$ from  Eq.~\eqref{eq:ptf} by a factor of $\approx 4$ to obtain the effective rms noise for point source detection: $\sigma_{\rm ps}=4\sigma$.  In our simulations, we conservatively assume $10\sigma_{\rm ps}$ threshold for detecting novae.

We parameterize our simulations via the nominal limiting magnitude ($m_{\rm lim}$) of the survey, which is related to the exposure time via the following for the sky limited case, typical of modern surveys 
\begin{equation}
m_{\rm lim}=ZP-2.5\times\log \left(\frac{5\sigma_{\rm sky}}{t_{\rm exp}}\right).
\end{equation}
Here $ZP$ is the zero point of the  photometric calibration and $\sigma_{\rm sky}$ is the noise from the sky background in a typical aperture for star flux measurement. The latter is obtained from Eq.~\eqref{eq:ptf} with $S=S_{\rm sky}$ and appropriate aperture correction as described above.

The intrinsic extinction map of M31  is computed based on the results of \citet{Tempel-2011} to which a constant foreground reddening  of $A_{\rm R}=0.15$ \citep{Shafter-2009} is added. The extinction is applied to all the seeded novae, depending on their spatial location in M31. 
However, its  overall impact  on the detection completeness does not exceed a few percent.

Taking into account the visibility of M31 from the northern hemisphere (between August and March) we assume the survey duration to be 211 days. 
For every decline time $t_2$ in the range of interest, we seed at every pixel of the SDDS image 20,000 novae occurring randomly in time within the survey time span. To each of the simulated novae, we assign a peak magnitude that has been drawn randomly from a Gaussian distribution with the mean and standard deviation as computed in  Appendix \ref{mmrd} (Fig.~\ref{fig:mmrd}, Table~\ref{table:mmrd}). The nova light curve is generated using the template derived in Appendix~\ref{template}, rescaled to have the desired $t_2$ and peak magnitude, Eq.~\eqref{eq:transformation} and then the extinction  is applied.
The detection efficiency  is determined for every pixel of the SDSS image  as the ratio of the number of detected novae, whose decline time $t_2$ could be measured, to the total number of seeded novae. 
To obtain the overall efficiency $\eta(t_{2})$, these values are then averaged across the image, with the weights proportional to the stellar mass contained in the given pixel (i.e., we assume that the nova rate scales with the stellar mass). To characterise the latter, we use the Spitzer 3.6~micron mosaic image of M31 by \citet{Barmby}. We thus compute the completeness curves for every survey configuration we have set up, determined by the observing pattern (every night, second night, third night, 75\% and 50\% random coverage) and limiting magnitude (19 to 22~mag).

Results of these simulations are plotted in Fig.~\ref{fig:mock}. As could be expected, the detection efficiency declines towards small $t_2$, because  novae with shorter $t_2$ times  fade away faster and therefore are less likely to be detected  than their longer lasting counterparts. On the other hand,  the detection efficiency also drops towards large $t_2$. This is caused by the combined effect of their lower peak magnitudes (Fig.~\ref{fig:mmrd}) and the finite observation time span. The latter is obviously determined by the survey duration. For example, a survey conducted in two consecutive years would have better detection efficiency for slow novae, than those shown in Fig.~\ref{fig:mock} (but same for short novae).

\begin{figure}
\includegraphics{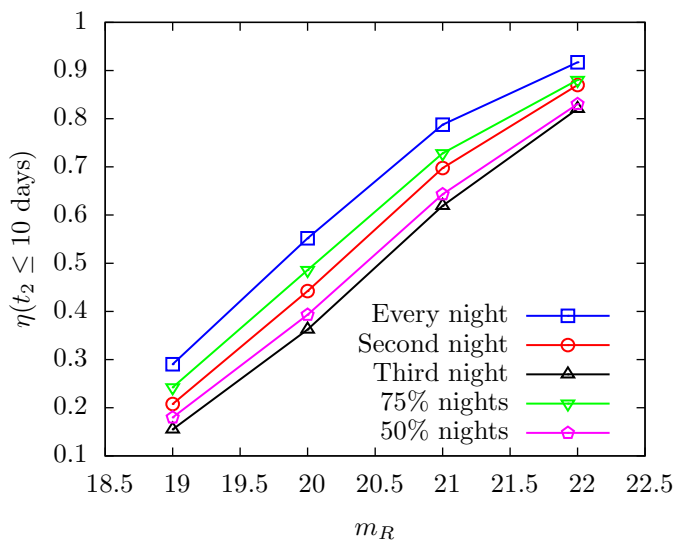}
\caption{ Fraction of all fast novae with $t_2\le10$~days detected in surveys of different temporal sampling as a function of their limiting magnitude. It is assumed that the distribution of novae over decay times $dn_{\rm nov}(t_2)/dt_2$ is given by Eq.~\eqref{eq:diff_rate}, for the mass accretion rate of $\dot{M}=10^{-7}~M_{\odot}~{\rm yr}^{-1}$.}
\label{fig:true_eff}
\end{figure}

The cumulative efficiency of the survey (i.e., the fraction of all novae with the $t_2$ time shorter than a given value, detected in the survey) depends on  $dn_{\rm nov}/dt_2$ -- the expected distribution of novae over $t_2$
\begin{equation}
\eta(\le t_2')=
\left( \int_0^{t_2'}\eta(t_{2})\frac{dn_{\rm nov}}{dt_2}dt_2 \right)
\left( \int_0^{t_2'}\frac{dn_{\rm nov}}{dt_2}dt_2 \right)^{-1}
\label{cumu_eta}
\end{equation}
The true distribution $dn_{\rm nov}/dt_2$  is unknown however. For example,  in the ``vanilla'' SD scenario from Sec.~\ref{sec:rate}, it is given by Eq.~\eqref{eq:diff_rate}. The result for $\dot{M}=10^{-7}~M_{\odot}{\rm yr}^{-1}$ is shown in Fig.~\ref{fig:true_eff} for surveys of different temporal sampling as a function of their limiting magnitude. As one can see from this plot, a high cadence survey with a limiting magnitude of $m_{R}=$~22 will detect about $\approx 80\mbox{--}90\%$ of fast novae ($t_{2}\le10$~days). In order to detect more than $50\%$ of  fast novae in a survey with observations conducted every night, its limiting magnitude has to be better than $m_{R}\approx$ 20. If observations are carried out every 3rd night, a limiting magnitude of $m_{R}\approx$ 20.5 is required.

To see how  the detection efficiency translates into absolute number of novae, let us consider the recent PTF survey of M31 as an example. PTF has been conducting regular and frequent M31 observations during the corresponding visibility periods (approximately from July/August to December/January). From Fig.~2 of \citet{Cao} we estimate that the observing schedule typically covers $\approx 50\mbox{--}80\%$ of nights during these periods every year. For the 5 sigma limiting magnitude of $m_{R}\approx20.6$ \citep{Law}, we estimate from Fig.~\ref{fig:true_eff} the detection efficiency of $\sim 55-65\%$. The ``vanilla'' SD scenario predicts about $\approx 3600$ fast ($t_2\le10$~days) novae per year in M31.  One should also take into account that the M31 visibility periods for the PTF telescope cover approximately 0.5 year. We therefore predict that the PTF survey of M31 should be detecting of the order of $\sim 1000\times f$ fast ($t_2\le10$~days) novae per observing season\footnote{\label{note:ptr}This estimate is valid for a random observing pattern. For the particular schedule of PTF observations in 2009--2010 shown in Fig.~2 of \citet{Cao}, it should be decreased by a factor of few, because of presence of extended gaps in the observing schedule,  more accurate calculations being beyond the scope of this work.}, where, as before,  $f$ is the fraction of SNe~Ia accreting in the unstable nuclear burning regime shortly before the explosion (at $M_{\rm WD}>1.3~M_\odot$). 
Furthermore, should all the observed POINT-AGAPE novae become SNe~Ia (i.e., $f=0.07$), a PTF program  with regular monitoring of M31 several times per week should be detecting of the order of $\sim 70$ fast novae per observing season.$^{\tiny \ref{note:ptr}}$ These numbers are significantly larger than the currently observed rate of fast novae, suggesting that $f\ll1$.

\section{Discussion}\label{disc2}

With the progenitors of SNe~Ia still having  eluded direct detection, there is a growing consensus that they may be a heterogenous class of objects united by the final outcome -- thermonuclear disruption of the WD. Among other possibilities, various  types  of accreting binary systems have been proposed as a candidate. The  fate of the accreted material is mainly determined by the WD mass and the mass accretion rate (e.g., \citealt{Fujimoto, Nomoto-2007, Wolf}). In the picture, which has become fairly standard, there is a rather narrow range of the mass accretion rates around $\sim{\rm few}\times 10^{-7}$~M$_\odot$~yr$^{-1}$ (varying with the WD mass) in which the nuclear burning is steady and proceeds at the rate determined by the supply of the material through the accretion process. Below this range, the nuclear burning is subject to thermal instability, giving rise  to the phenomenon of novae. In the classical  picture, copious amounts of material are lost in the nova explosion, especially at the lower mass accretion rates ($\dot{M}\la 10^{-7}-10^{-8}$~M$_\odot$~yr$^{-1}$ in the \citet{Yaron} calculations for example), rendering the growth of the WD mass impossible or insignificant in the context of SN~Ia progenitors. Importantly, several authors have found that  the transition from unstable to stable burning is sharp -- there is a discontinuity in the stability of burning, with the large amplitude flashes occurring very close to the stability strip (e.g, \citealt{Wolf, Kato-2014}).

However, the main aspects of this picture have been contested by several authors. Firstly, existence of the stability strip has been questioned (e.g., \citealt{Idan-2013, Starrfield-2014, Hillman-2015}). In fact, no stable burning is reported in the \citet{Yaron} tables either, although the ejected mass becomes zero at the largest mass accretion rates.
Secondly, it has been suggested that mass accumulation at appreciable rates is possible in the broad range of the mass accretion rates, including those traditionally considered to be associated with the nova regime. In fact, it has been claimed  that any mass accumulation by the WD is associated with regular  thermonuclear explosions \citep[e.g.,][]{Starrfield-2014, Hillman-2015}. In a related development, progenitor models have been proposed, which involve WDs accreting in the presumably unstable nuclear burning regime throughout or at least a part of their accretion history (e.g., \citealt{Starrfield-1985, Hachisu-2001, Hillman-2015}). Further support to these ideas is lent by the realisation that  recurrent novae (RNe) may host WDs with the mass very close to the Chandrasekhar mass limit, the most famous example being RS Oph \citep{Sokoloski, Hachisu-2001}. This led to the suggestion that novae in general and recurrent novae in particular may be an important channel  producing some (unknown) fraction of SNe~Ia (see, for example, \citealt{Vasey-2006,Patat-2011, Hachisu-2001, Pagnotta-2014}).

In this paper, we point out that the population of WDs with unstable nuclear burning, sufficient to account for a non-negligible fraction of SNe~Ia, would reveal themselves through significantly enhanced nova rates in galaxies. We thus propose that  the contribution of novae to the observed SN~Ia rates can be assessed through the nova statistics in nearby galaxies.  We demonstrated that given the completeness corrected nova rate in M31 of $\approx 65$~yr$^{-1}$ \citep{Darnley}, novae can only produce a small fraction of SNe~Ia, at the maximal rate of $\la (1$--$5)\times 10^{-4}$~yr$^{-1}$, assuming typical mass accretion rates of progenitors in the range, $\dot{M}\sim(1$--$5)\times10^{-7}$~M$_\odot$~yr$^{-1}$. This constitutes no more than $2\mbox{--}7\%$ of the total SN~Ia rate in M31. Moreover, we predicted that M31 surveys of the PTF class should be detecting of the order of $1000\times f$ fast ($t_2\le10$~days) novae every year, where $f$ is the fraction of SNe~Ia accreting in the unstable nuclear burning regime shortly before the explosion (at $M>1.3~M_\odot$). The fact that $< 50$ such fast novae have been recorded/detected in the course of about a century of M31 monitoring (see the optical nova catalog maintained by MPE from Footnote~\ref{MPE_table}) suggests that the fraction $f$ is small, of the order of $f\la 10^{-3}$. However, an accurate completeness analysis of the M31 PTF survey is needed before a robust quantitative conclusion can be made.

\begin{figure}
\includegraphics{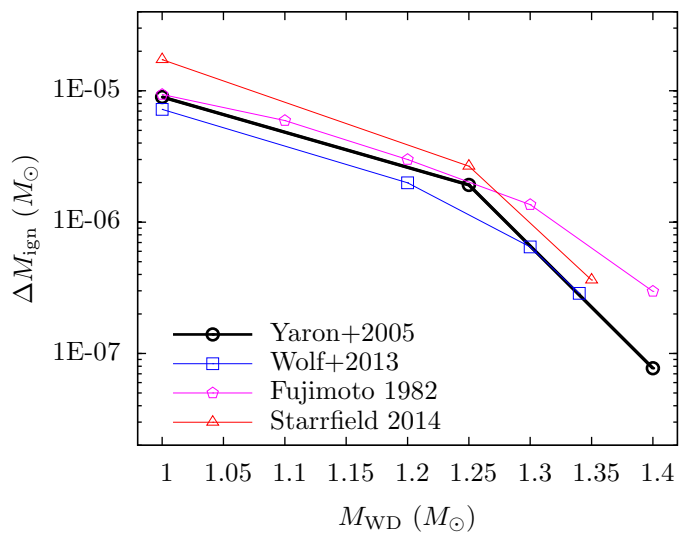}
\caption{Comparison of the ignition mass of the nova as computed by different authors. The ignition mass is shown as a function of the WD mass, for the mass accretion rate $\dot{M}=10^{-7}$~M$_\odot$~yr$^{-1}$. The \cite{Wolf} results are obtained from their Fig.~8; \citet{Fujimoto} from their Fig.~7; \citet{Starrfield-2014} from their Figs.~2--3.
\label{fig:mign}
}
\end{figure}

As one can see from Eq.~\eqref{total_rate}, the calculation of the total number of novae per SN~Ia depends on the theory of thermonuclear burning on the WD surface only through the ignition mass $\Delta M_{\rm ign}$. This quantity is derived from sufficiently well understood physical  principles and has been computed  by a number of groups (for example, \citealt{Fujimoto, Townsley, Yaron, Wolf, Kato-2014} and others).  Comparison of some of the results is shown in Fig.~\ref{fig:mign}. As one can see from the plot, results of different calculations agree quite well, within a factor of $\sim 2$ or so. The agreement is quite good even between authors coming to opposite  conclusions regarding the existence of the stability strip. In the parameter range of interest, there is a good agreement between analytical \citep{Fujimoto} and more sophisticated numerical calculations, as well as between single flash \citep{Fujimoto, Starrfield-2014} and multicycle \citep{Yaron, Wolf} calculations. We note that the latter two should provide a more accurate representation of the accreting WD; and agreement between their results is much better than a factor of $\sim 2$. We thus  conclude that the calculations of the total number of novae produced by an SD SN~Ia progenitor presented in this paper are sufficiently robust and do not depend on the details of the underlying nova models, including their conclusions regarding the existence of the stability strip.

\begin{figure}
\includegraphics{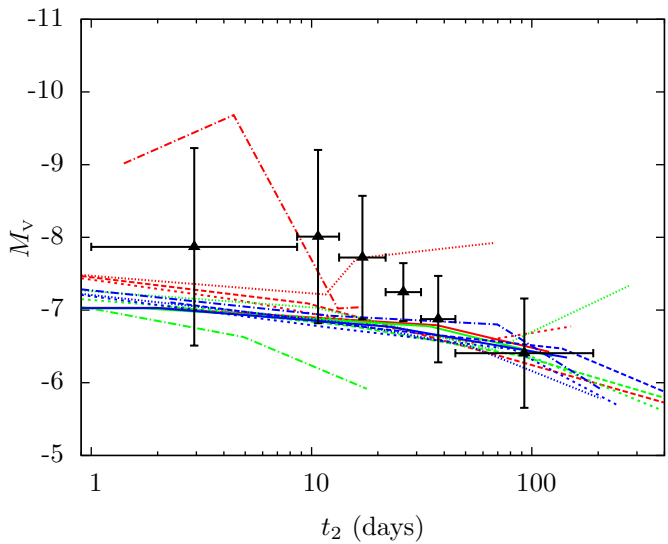}
\caption{Comparison of the data shown in Fig.~\ref{fig:mmrd} with the results of multicycle nova models of \citet{Yaron}. The black triangles  and vertical error bars show the average peak magnitudes and their rms deviation for the novae grouped in equi-populated bins over the $t_2$ time, and the curves are corresponding relations for the nova models of \citet{Yaron}. The red, green and blue curves correspond to WD core temperature of $10^7$~K, $3\times10^7$~K and $5\times10^7$~K, respectively. The solid curves are for $\dot{M}=10^{-7}~M_{\odot}~{\rm yr}^{-1}$, dashed curves for $\dot{M}=10^{-8}~M_{\odot}~{\rm yr}^{-1}$, short dashed curves for $\dot{M}=10^{-9}~M_{\odot}~{\rm yr}^{-1}$, dotted curves for $\dot{M}=10^{-10}~M_{\odot}~{\rm yr}^{-1}$ and dot-dashed curves for $\dot{M}=10^{-11}~M_{\odot}~{\rm yr}^{-1}$.}\label{fig:mmrd_yaron}
\end{figure}

In our calculations we used the results of \citet{Prialnik} and \citet{Yaron}, who computed the most extensive grid of multicycle nova evolutionary models to date. The ignition masses from their calculations agree well with the results of other groups, as already discussed above (Fig.~\ref{fig:mign}). Furthermore, the grid of their models covers well the parameter space occupied by the observed novae and approximately reproduce observed correlations between various nova parameters (\citealt{Prialnik}; see also \citealt{Walder-2008}). The peak magnitudes and light curve decay times predicted by their models are compared with observations in Fig.~\ref{fig:mmrd_yaron}. Each curve in this plot corresponds to a given combination of $\dot{M}$ and $T_{\rm WD}$, with $M_{\rm WD}$ changing along the curve. $\dot{M}$ extends from $10^{-7}$ to $10^{-11}~M_{\odot}~{\rm yr}^{-1}$, thus covering the typical range expected in nova host systems. Peak magnitudes in the visual band are computed from the peak bolometric luminosity reported in \citet{Prialnik} and \citet{Yaron} by applying the bolometric correction of an A5V star, as is commonly  done in such estimates \citep{Shafter-2009, Kasliwal}. The light curve decay times are computed from the mass-loss times as explained in Sec.~\ref{sec:rate}. 

As one can see from Fig.~\ref{fig:mmrd_yaron}, the observed range of the light curve decay times $t_2$ is fully sampled by the nova models with the WD mass varying from $\approx 0.4$--$0.65$~M$_\odot$ to  $\approx 1.3$--$1.4$~M$_\odot$.
The models also reproduce rather well the average trend in the peak magnitude with $t_2$.  The small offset in magnitude between the models and data is reduced further if one takes into account that the theoretical models may underestimate the true value by upto 0.75~mag, as discussed in \citet{Prialnik}. These models, however, do not reproduce the scatter seen in the observed data. This suggests that the scatter may be caused by additional factors, other than those already included in the calculations of \citet{Yaron} (e.g., orientation effects, interaction of the ejecta with the disk and the donor star, etc.). For this reason we used the parameters of observed novae  to draw the nova peak magnitudes in our Monte-Carlo simulations in Sec.~\ref{incompleteness}.
However, the overall agreement between the \citet{Yaron} models and the data suggests that the models reproduce the global characteristics of the nova light curves sufficiently well.

The famous recurrent nova M31N~2008-12a appears to be somewhat (by about $\sim 1\mbox{--}2~{\rm mag}$) underluminous as compared to the \citet{Yaron} predictions as well as to  its counterparts with similarly short decay time (Fig. \ref{fig:mmrd}). Of course, it may be just a faint tail of the distribution approximately centred near the value predicted by  \citet{Yaron} for the corresponding range of decay times. However, there is another possibility that it presents the {\em bright} tail of the so far unknown population of fast and underluminous novae hosted by massive WDs near the Chandrasekhar mass limit. Although intriguing, this possibility seems at present less likely. The main argument is that M31N~2008-12a is up to 3 magnitudes brighter than the typical sensitivity limit of the PTF M31 survey, and despite around 6 years of continuous PTF observations (since August 2009) no other similar nova was discovered. More quantitative analysis of the latter possibility is beyond the scope of this paper.

\citet{Shafter-2015} have undertaken a census of the recurrent nova population in M31 based on the positional coincidence of about $\sim 10^3$ novae recorded in modern astronomy. They identified 16 recurrent novae and candidates, however they estimated that the detection efficiency of the recurrent systems may be as low as $\sim 10\%$ of that of classical novae.  Based on these data and assuming that recurrent nova systems typically accrete at $10^{-7}~M_{\odot}~{\rm yr}^{-1}$, they constrained their contribution to the SN~Ia rate at the level of $\la 2\%$. This upper limit is comparable to the limits derived in this paper. In particular, for the $10^{-7}~M_{\odot}~{\rm yr}^{-1}$ accretion rate, we constrained the contribution of all slow ($t_2\ga 10$~days) novae, irrespective of their recurrence times, to $\la 7\%$.

The upper limit of $\approx 2\mbox{--}7\%$ obtained in this paper is comparable to the upper limits on other versions of the SD scenario. Based on the luminosity of unresolved soft X-ray emission  in  a sample of nearby elliptical  galaxies observed by {\em Chandra}, \citet{Gilfanov-2010} constrained the contribution of supersoft X-ray sources (i.e., stably nuclear burning WDs located in the stability strip) to $\la 5\%$.   \citet{Johansson} used the recombination line $\lambda4686$~{\AA} of HeII to  limit the EUV emission from lower temperature sources -- those that escaped the X-ray based analysis of \citet{Gilfanov-2010} due to their very soft spectra and absorption by the ISM. Such low temperature sources can be associated with the low mass WDs or with the rapidly accreting sources above the stability strip \citep{Hachisu-1996}.  In particular, \citet{Johansson} applied the diagnostics suggested by \citet{Woods-a} to  stacked SDSS spectra of $\sim 10^4$ retired galaxies, and constrained the contribution of accreting WDs with photospheric temperature in the range $\sim (1.5\mbox{--}6)\times 10^5$~K to $\la 5\mbox{--}10\%$. These upper limits can be tightened further and their photospheric temperature range can be extended  using intrinsically brighter forbidden lines of metals \citep{Woods-b}. In particular, Johansson et al. (in preparation) used the $\lambda6300$~{\AA} forbidden line of neutral oxygen to derive a $\sim $few per cent upper limit on the contribution of accreting WDs with the photospheric temperature in the $\sim 10^5\mbox{--}10^6$~K range. Combined together, these constraints limit the role of the main variations of the SD scenario in producing SNe~Ia. Even when (very conservatively) summed up independently, their total contribution to the observed SN~Ia rate cannot exceed $\la 10\mbox{--}20\%$. Alternatively, a typical supernova could not have accreted more than $\la 0.03-0.05$~M$_\odot$ in each of the above mentioned regimes -- i.e. below, in or above the stability strip. 

We assumed, as it is commonly accepted \citep[e.g.,][]{Maoz}, that SNe~Ia are produced by the WDs exploding at the Chandrasekhar mass. However, sub-Chandrasekhar (e.g., \citealt{Woosley-1994, Bildsten-2007, Sim-2010, Kromer-2010}) as well as super-Chandrasekhar (e.g., \citealt{Howell-2006, Liu-2010, Kamiya-2012}) detonations  are also being considered by a number of authors.  If sub-Chandrasekhar detonations contribute significantly to the SN~Ia rate, the nova-based constraints may need to be relaxed. Indeed, if a  WD explodes before reaching the Chandrasekhar mass, the distribution shown in the right panel of Fig.~\ref{fig:cum_diff} will be cut-off at the $t_2$ time, corresponding to the explosion mass (cf.~Fig.~\ref{fig:ignition}). Correspondingly, the upper limit on the contribution of novae to SN~Ia production will increase, as $n_{\rm nov}$ in Eq.~\eqref{eq:max_sn_rate} decreases. For example, if a typical SN~Ia progenitor WD explodes at $1.3~M_\odot$, the entire population of fast novae ($t_2<10$ days) will not be produced by the sub-Chandrasekhar supernova progenitors, but the constraints derived form the POINT-AGAPE survey will still hold in full. Obviously, in this scenario the observed fast novae are not the progenitors of sub-Chandrasekhar supernovae.  If the typical explosion mass is yet lower, for example $1.1~M_\odot$,  current nova statistics become unconstraining. In the super-Chandrasekhar models the result depends on the properties of nova explosions on the surface of a rotating WD, which are currently not well understood.  However, at present there is no evidence that either sub- or super-Chandrasekhar models make dominant contribution to the SN~Ia rates, therefore constraints on the nova-channel derived in this paper should hold for the bulk of supernovae.

Finally, we note that we used M31 galaxy to tune our calculations and compare their results with observations, due to its proximity and relatively well known nova population. However, our results can be easily generalised to other galaxies of similar age. Indeed, observations of several nearby galaxies show that they have comparable nova rates per unit stellar mass (e.g., \citealt{Shafter-2014}). As same is true for the supernova rates in galaxies of similar morphological type, the constraints on the contribution of the nova channel to supernova rate should apply to other nearby galaxies of similar age and morphological type.

\section{Conclusions}\label{conclusion}

We propose that the statistics of novae in nearby galaxies is a sensitive  diagnostic of the population of accreting WDs with unstable nuclear burning, and can be used to constrain the role of this channel in producing SNe~Ia. Using multicycle nova  models of \citet{Yaron}, we compute the number and temporal distribution of novae produced by a successful SN~Ia progenitor, assuming that it accretes in the unstable nuclear burning regime throughout its accretion history.  We predict the total number of novae of  $\approx 6.5\mbox{--}7.7\times 10^{5}$ per supernova, assuming that a typical SN~Ia progenitor accretes material at the rate of $\approx (1$--$5)\times10^{-7}$~M$_\odot$~yr$^{-1}$. Using the   nova rate in M31 as measured by  the POINT-AGAPE survey, $\approx 65$~yr$^{-1}$ \citep{Darnley-2004, Darnley}, we estimate the maximal contribution of the nova channel in producing SNe~Ia.  Considering relatively slow novae, with light curve decay times $t_2\ga 10$~days, whose characteristics are compatible with the observed novae from the POINT-AGAPE catalog \citep{Darnley},  we conclude that their contribution to the observed SN~Ia rates cannot exceed $\approx (1-5)\times 10^{-4}~{\rm yr}^{-1}$. This constitutes less than 2--7\% of the total SN~Ia rate in M31.

An even more sensitive diagnostic can be provided by fast novae, which are the ones originating on the most massive WDs, characterised by the smallest ignition masses.  To utilise their potential, high cadence nova surveys are required. We investigate how detection efficiency of a generic nova survey of M31 depends on its limiting magnitude and temporal sampling.
We find that a survey with a limiting magnitude of $m_{R}\approx 22$ will detect about $\approx 80$--$90\%$ of the predicted fast novae ($t_2\la 10$~days) provided that observations are conducted at least every 2nd or 3rd night.  In order to detect more than $50\%$ of such novae in a survey with observations carried out every night, its limiting magnitude has to be $m_{R}\gtrsim$20. 
Such surveys should be detecting of the order of $\gtrsim 1000\times f$ fast novae per observing season in M31, where $f$ is the fraction of SN~Ia progenitors which accreted  in the unstable nuclear burning regime while accumulating the final $\Delta M\approx0.1~M_\odot$  before the supernova explosion. This  is significantly larger than the currently observed rate of fast novae in M31, suggesting that $f\ll1$. However, high cadence surveys with accurately characterised completeness are required to place robust constraints on the value of $f$.
The existing and upcoming surveys like the PTF, Pan-STARRS and  LSST are  well-suited for this task.

The predicted number of novae per SN~Ia should not  depend strongly on the details of the underlying nova models, as they are determined  only by the  ignition mass of the envelope. This quantity is derived from sufficiently well understood physical principles and results of computations by different groups agree quite well, even among those at odds about the existence of the ``stability strip''.  The division of the total number between ``slow'' and ``fast'' novae depends on the assumption about the shape of their light  curves. To this end, we used the multicycle nova  models of \citet{Yaron}, which are known to  sample correctly the observed range of the nova decay times and reproduce various correlations between observed nova properties. Calculations based on two different timescales tabulated in \citet{Yaron} -- mass-loss and bolometric, give similar results leading to the same conclusions. Finally, in our incompleteness simulations  we used the observed nova light curves and their peak magnitudes,  therefore these results are independent of the theoretical nova models.

\section*{Acknowledgments}
We would like to thank Dr. Chien-Hsiu Lee for making available to us the tabulated data of the detected novae from WeCAPP. We are grateful to Dr.~Pauline~Barmby for providing us the Spitzer 3.6~micron mosaic image of M31. MG acknowledges hospitality of the Kazan Federal University (KFU) and support by the Russian Government Program of Competitive Growth of KFU. The authors would like to  thank the anonymous referee for the constructive and inspiring  comments and suggestions which helped to improve the paper.

\bibliographystyle{aa}
\bibliography{ref}

\appendix

\section{The nova light curve template }\label{template}

\begin{figure*}
\vspace{4mm}
\includegraphics{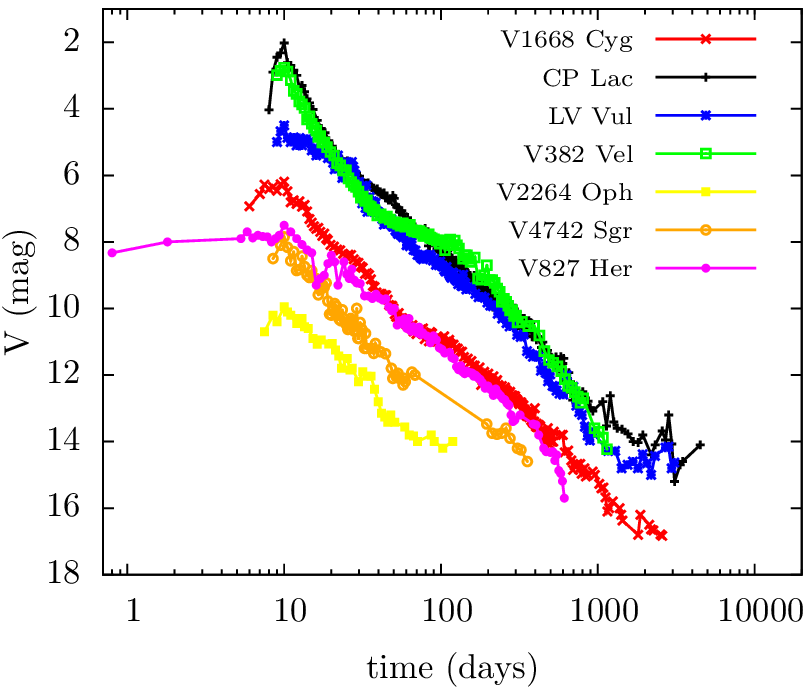} \hfill \includegraphics{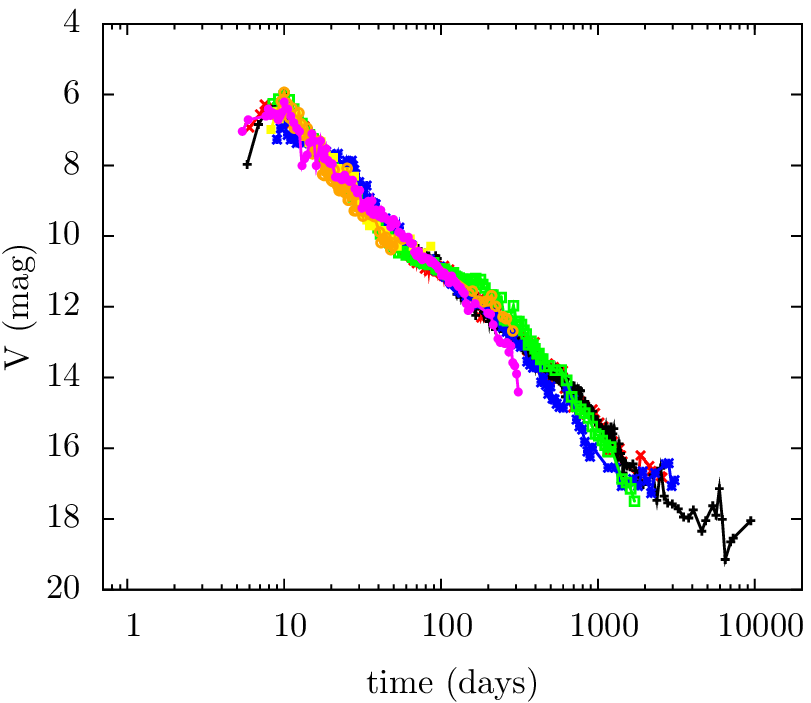}
\caption{ Light curves of the sample selected from the S class of the catalog compiled by \citet{Strope} with all the peaks aligned at $t=10$ days by a linear shift along the time axis ($left$) and the resulting transformed curves after stretching in time and shifting in magnitude with respect to the reference light curve of V1668 Cyg according to Eq.~\eqref{eq:transformation} ($right$).}\label{fig:strope}
\end{figure*}

Nova light curves are known to have a variety of  shapes. In an effort to classify them, \citet{Strope}, using a  sample of 93 well-observed Galactic novae, have proposed seven morphological classes. 
Of these, the ``smooth'' (S) class is the most fundamental since the other light curve shapes  can be derived from the smooth class  by superposing it with various features. The largest fraction of novae in the \citet{Strope} sample ($38\%$)  falls in the S class; more than half of the novae with $t_{3}\leq21$~days (taking this value to define the fast novae of interest) belong to this class as well. Furthermore, in our simulations, we will be using light curves within 2--3~magnitudes from the peak, where they  generally have smooth morphology (most of the features defining the distinction between the various classes typically develop much later in time). 
We have, thus, adopted the S class light curves for generating the template curve as described below.

We select all sufficiently well-sampled S class light curves from \citet{Strope}. In Fig.~\ref{fig:strope}, these light curves are shown with their peaks aligned by shifting linearly along the time axis. We then use the following transformation to match the light curves: 
\begin{equation}
\label{eq:transformation}
m(t)\rightarrow m'(t')=\Delta{m}+m[(t-t_{p})s],
\end{equation}
where $m(t)$ is the magnitude in the frame $t$, $m'(t')$ is that in the transformed frame $t'=(t-t_{p})s$, $t_{p}$ is the time of the peak of the light curve, $\Delta{m}$ the magnitude shift and $s$ the time stretch factor. 

The light curves are transformed to match the reference light curve, for which we have chosen V1668 Cyg. To this end, we bin the light curves logarithmically (typically 5 bins or more per dex along the time axis) and determine the magnitude in each bin by taking the average, weighted by the inverse square of the uncertainty in the individual magnitude measurement. The best fit parameters  $\Delta{m}$ and $s$ of the transformation Eq.~\eqref{eq:transformation} are determined by minimizing the $\chi^2$.
The result of this procedure is shown in the right panel in Fig.~\ref{fig:strope}. In all cases we were able to obtain a reasonably good agreement, with the values of the stretch factor ranging from $\approx0.5$--2 and the rms dispersion between the reference and individual light curves calculated down to 6~mag from the peak being in the range 0.18--0.33~mag. Finally, we average the resulting transformed curves by binning in time (logarithmically again) and weighting by the respective uncertainties to produce the template curve.

\begin{figure}
\includegraphics{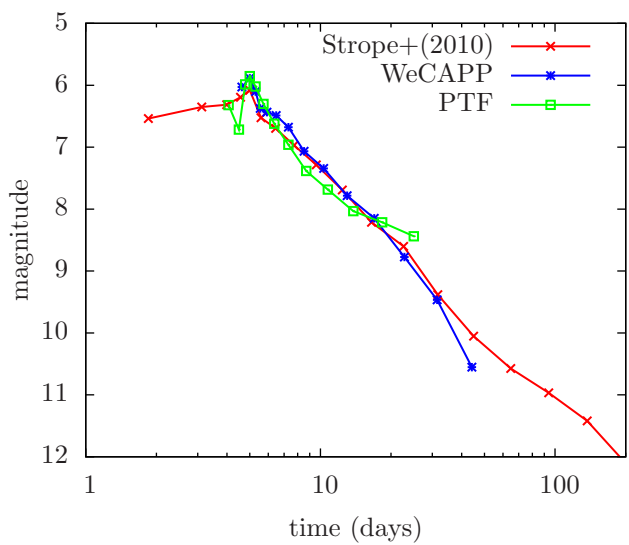}
\caption{Co-aligned template light curves from the three nova samples, viz., the Galactic sample from \citet{Strope} and the M31 samples obtained from the WeCAPP and PTF nova catalogs.}\label{com_templ}
\end{figure}

The above template light curve is generated using a Galactic nova sample. Although we do not expect any significant difference from the M31 novae, for a consistency check we compare our template with light curves from  two  different  M31 nova samples. We apply the  procedure described above to the selection of well-sampled light curves from the nova catalog of the WeCAPP \citep{Riffeser}, which have been classified as having smooth morphology \citep{Lee}, and to light curves from the PTF M31 survey in \citet{Cao}. In these two cases too, we were able to obtain good light curve matches, with the rms dispersion (calculated down to 4~mag from peak) less than $\approx 0.30$~mag. Final template curves are generated using the same procedure as before. The resulting templates  are shown in Fig.~\ref{com_templ} along with our default template based on \citet{Strope} data. As is obvious from the plot, all three templates agree very well with each other. For our calculations, we use the template obtained based on the  \citet{Strope} data as it is the best sampled and covers the broadest magnitude range.

\section{Peak magnitudes of novae}\label{mmrd}

For each simulated nova we need to assign some value for its peak magnitude. To this end, we produce an analog of the MMRD relation with the caveats detailed below. 
Various forms of this relation  have been obtained previously (for example, see \citealt{Della, Downes, Darnley, Shafter-2011}), although recently its existence has been questioned \citep{Kasliwal}. However, for the purpose of this study, we are not concerned with the existence of the MMRD relation, instead we need to account for the full range of observed nova peak magnitudes in our simulations.
We therefore collected a large  sample of novae and determined their mean magnitudes and rms scatter in broad bins over $t_2$ as explained in the following.

We used the light curves from  the PTF \citep{Cao} and WeCAPP \citep{Lee} nova catalogs with morphological classification available, the light curves from \citet{Shafter-2011} that have been observed in the V-band, the high-quality light curves from \citet{Capaccioli} and the sample of extragalactic novae discovered by P60-FasTING \citep{Kasliwal}. To this compilation, we added the recently discovered very fast recurrent nova M31N~2008-12a in M31 with  the shortest known recurrence period of $\sim 1$~year \citep{Shafter-2012, Darnley-2014, Henze-2014, Tang}. The PTF and WeCAPP light curves were converted from  R-band to V-band using the colour $(V-R)_\circ=0.16$ \citep{Shafter-2009} after accounting for the foreground extinction of $A_{\rm R}=0.15$ \citep{Shafter-2009} estimated using a reddening of $E(B-V)=0.062$ along the line of sight to M31 from \citet{Schlegel}. The light curves from \citet{Capaccioli} were  corrected for foreground extinction using $A_{\rm pg}=0.25$ \citep{Shafter-2009} and converted from the photographic band to V-band using the colours $(B-m_{\rm pg})_\circ=0.17$ \citep{Capaccioli, Arp} and $(B-V)_\circ=0.15$ \citep{Shafter-2009}. The  \citet{Kasliwal} light curves were converted from  {\it g}-band to V-band using the transformation relation from \citet{Jordi}.  Apparent magnitudes were converted to absolute magnitudes using a distance modulus of 24.36 for M31 \citep{Vilardell}. Light curves from \citet{Shafter-2011} were corrected for extinction and  converted to absolute magnitude in the original publication.  Except for the WeCAPP sample, we use the light curve decline times ($t_2$) from the respective publications. For those novae in \citet{Kasliwal} whose $t_2$ times are not available, we approximate them by multiplying the decline times $t_1$ in their Table~5 by two. For the WeCAPP light curves we estimate their decline times ourselves, by linearly interpolating between consecutive measurements.

\begin{figure}
\includegraphics{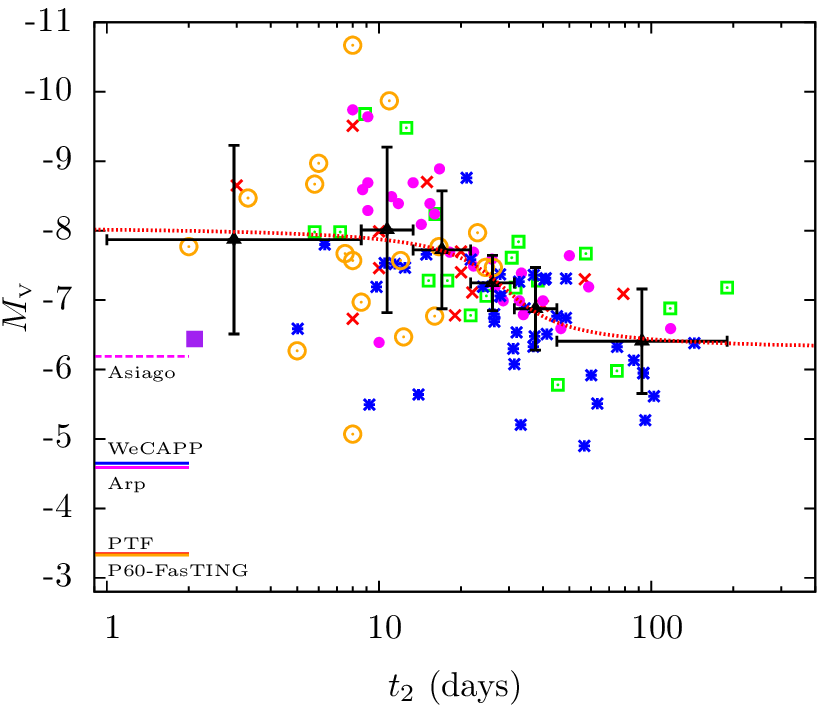}
\caption{Relation between the maximum magnitude and the rate of decline for extragalactic novae. The  PTF data are shown by  red ``x'' symbols, WeCAPP -- blue stars,  \citet{Shafter-2011} data -- green squares, \citet{Capaccioli} data --  magenta filled circles, \citet{Kasliwal} data  --  orange open circles and the purple square is the recurrent nova M31N~2008-12a in M31. The black triangles represent the averages in the bins and the red dotted line is our fit. The line segments on the left side mark the approximate limiting magnitudes of the surveys (as shown by the legend).}\label{fig:mmrd}
\end{figure}

\begin{table}[h]
\renewcommand*{\arraystretch}{1.2}
\caption{Mean nova peak magnitudes and their standard deviations}
\label{table:mmrd}
\begin{tabularx}{250pt}{X X c}
\hline
$t_{2}$~range	&$<M_{\rm V}>$	&$\sigma$\\
(days)	&  (mag)	&(mag)\\
\hline
$<8.60$	&$-7.87$	&1.36\\
$~8.60 - 13.33$	&$-8.01$	&1.19\\
$13.33 - 21.70$	&$-7.72$	&0.85\\
$21.70 - 31.41$	&$-7.25$	&0.40\\
$31.41 - 44.98$	&$-6.88$	&0.59\\
$>44.98$ 	&$-6.41$	&0.75\\
\hline
\end{tabularx} 
\end{table}

The collected data  are shown in Fig.~\ref{fig:mmrd}. In order to quantify the data in a statistical manner, we grouped them according to the decline times, with the bin-width adjusted to have equal number (19) of novae in each bin. For each bin we computed  the mean magnitude and its standard deviation (Table~\ref{table:mmrd}), the values of which are fed into our simulations as described in Sec.~\ref{incompleteness}.

As a final word of caution, the data used in computing Table~\ref{table:mmrd} are collected from a heterogeneous set of magnitude limited surveys. Therefore, our results should not be interpreted as an attempt to produce an updated version of  the  MMRD relation. Incompleteness of individual surveys could bias our calculations of mean magnitudes, shifting them upwards. 
However, the magnitude limits of the majority of the surveys used here are  deep enough (Fig.~\ref{fig:mmrd}), therefore our results should be sufficiently accurate for the purpose of an illustrative calculation undertaken here.

\end{document}